# Al-doped ZnO aligned nanorod arrays: Significant implications for optic and opto-electronic Applications

T. Holloway, R. Mundle, H. Dondapati, M. Bahoura, A. K. Pradhan[a]

[a]Center for Materials Research, Norfolk State University, 700 Park Ave., Norfolk, VA 23504, USA
apradhan@nsu.edu

**Abstract.** We investigated the optical and optoelectronic properties of vertically aligned Al:ZnO nanorod arrays synthesized by the hydrothermal technique at considerably low temperature on a sputtered Al:ZnO seed layer. The nanorod arrays maintain remarkable alignment along to the c-axis over a large area. The seed layers as well as the nanorod arrays showed various optical band gaps. Investigation of the optoelectronic properties of nanorod arrays on Al:ZnO/p-Si seed layer with $SiO_2$ showed that the photocurrent is significantly reduced in nanorod arrays on $AZO/SiO_2/p$-Si heterojunction due to multiple scattering phenomena associated with the nanorod arrays. This research may open up venues for various optical and opto-electronic applications where highly aligned nanostructures are desired.

## 1 INTRODUCTION

Vertically aligned nanorod arrays provide a simple matrix to study the average effect of assembled nanorods [1-4]. One-dimensional (1-D) semiconductor nanomaterials have been attracting increasing attention due to their outstanding properties, which are different from bulk materials. Particularly, well-aligned ZnO nanorod arrays show great potential for solar cell applications [5-6]. Nanorods have been used to increase pn-junction area in dye-sensitized solar cells [7] and in polymer semiconductor hybrid solar cells [8]. High mobility nanorods are pathways for exciton diffusion into pn-junctions in solar cells [9]. The charge transport as well as tailoring of band gap by either doping or quantum confinement due to size effects are considered to be very import issues for the application of semiconductor nanostructures, especially for solar cell applications. Furthermore, template-free (for example, not using porous alumina template) well aligned growth of nanostructure on the microstructural and optical characteristics of the solution- grown Al-doped ZnO nanorod arrays have significant impact for optical, electronic and opto-electronic applications. Zinc oxide is a direct n-type semiconductor with a wurtzite hexagonal structure, presenting a wide band gap (3.3 eV at room temperature) and a large exciton binding energy (60 meV). Highly oriented vertical nanorod arrays are highly desired for their future application in various fields like nanoelectronic or photoelectronic materials where 1D directionality is needed (lasing, photonic effects, etc). Also, light scattering properties are under intense focus for application in the field of solar cells in order to enhance the absorption of light in thin absorber layers.

In this paper, we report almost perfectly vertically aligned ZnO nanorod arrays synthesized by the hydrothermal route at 90 °C on a sputtered Al:ZnO seed layer. The nanorod arrays demonstrate remarkable alignment along to the c-axis over a large area. The optical properties of the seed layers as well as the nanorod arrays were investigated. The optoelectronic properties of nanorod arrays on Al:ZnO/p-Si seed layer with $SiO_2$ have been illustrated. Our results clearly demonstrate that the photocurrent is significantly reduced in nanorod arrays on AZO/SiO$_2$/*p*-Si heterojunction due to multiple scattering phenomena associated with the nanorod arrays. However, this can be an opportunity for reabsorption of light by dye or photovoltaic polymers for enhanced effects.

## 2 SAMPLES PREPARATION

Al: ZnO (AZO) seed layer films were deposited to Boron doped p-Si and glass substrates by rf sputtering technique using an Ar pressure of 2.2 mTorr, rf power, 150 W, and the deposition temperature, 350°C for deposition time of 20-30 minutes. An AZO (2 wt·% Al$_2$O$_3$) ceramic target was used to deposit the Al:ZnO seed layer film. The Si substrate was cleaned using a mixture of HF:H$_2$O (1:10) in order to remove SiO$_2$ and achieve H-terminated surface.

AZO nanorod arrays were grown by the hydrothermal method [8]. The AZO seed coated samples were placed in an aqueous solution (50ml) of different molarity concentration of 0.1M, 0.05M zinc nitrate (Zn(NO$_3$)$_2$), 0.1M, 0.05M hexamethylenetetramine (HMT), and different wt.% (1%, 5%, 10%) of Al doping. For Al doping, Al(NO$_3$)$_2$ was used to explore the effect of each chemical on the growth parameters of the nanorod arrays. All precursors were dissolved in deionized water. A glass bottle with the cap screwed on tightly was used and heated in a silicon oil bath at a 95ºC, for 2 to 4 hours. At the end of the growth period, the samples were cooled to room temperature and removed from the solution. Then the samples were immediately rinsed with deionized water to remove any residual salt from the surface. Finally, the samples were blown with nitrogen and dried in air in an oven at 100ºC for 10-15 min. It is noted that we concentrate on for only concentration, ie., 5 wt% of Al doping by nitrate precursors.

Thickness measurement of the seed layer was performed using Dektak porofilometer, and found to be about 150 nm. The XRD measurements were performed on a Rigaku powder X-ray diffractometer with Copper K$\alpha$ incident radiation ($\lambda$=1.5405 Å) to examine the crystallographic structure. X-ray diffraction studies show intense (002) peak, indicating Wurtzite structure and most of the nanorods grow with their c-axis normal to the surface, indicating that the nanorod arrays aligned vertically to the surface of the seed layer . In further examination of the XRD curves, a gradual increase in the (100), (002), and (101) peak intensity was seen with annealing temperature, indicative of improved crystallinity.

Transmission Electron Microscope (TEM) was used to determine the crystal structure and morphology, Field Emission-Scanning Electron Microscope (FE-SEM), Hitachi 4700 Scanning Electron Microscope, as well as Scanning Electron Microscope (SEM), JEOL JSM-5900LV, to examine the surface morphology and orientation of the nanorods arrays. The Atomic Force Microscopy (AFM) study of the surface roughness and surface grain size of ZnO seed layers was performed using a Bruker dimension Icon AFM microscope. The absorption was measured by UV-VIS-IR spectrophotometer (Perkin Elmer Lambda 950 UV-VIS-IR). The optical absorption was measured against a bare glass as a reference to obtain the absorption of the AZO seed layer. Air and AZO seed layer were used as a reference sample to obtain absorption of the AZO nanorod arrays. Temperature dependence of the electrical resistivity was measured using a four-probe technique. The photocurrent was measured from the current-voltage characteristics using a solar simulator.

## 3 RESULTS AND DISCUSSION

Figure 1 shows the typical FE-SEM and AFM images of the seed layers. The SEM image (Fig. 1 (a) of the seed layer clearly illustrate that the grains are very uniform in size. The grain size is typically 30 nm in size. This is being confirmed by the AFM image shown in Fig. 1 (b). The surface roughness is about 1nm. The films are optically very smooth with having any microscopic defects. The major merit of using the seed layer is to nucleate and guide the nanorod growth in an array form. On the other hand, it provides template free growth opportunity with electrical contact and optical transparency of the seed layer, because AZO is a transparent conducting oxide (TCO).

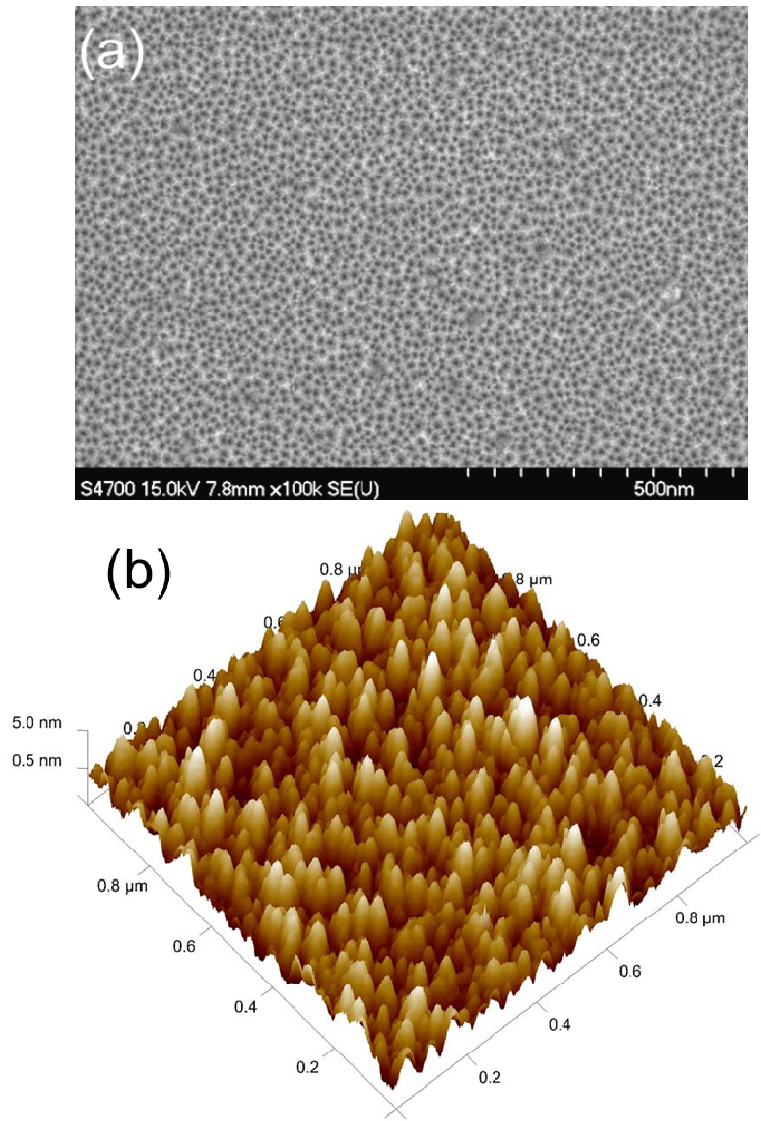

Fig. 1. (a) FE-SEM image of Al:ZnO seed layer and (b) is AFM image of the seed layer (1 μm × 1μm).

Figure 2 (a) - (c) shows the SEM images of Al:ZnO nanorod arrays with 5 wt.% of Al in ZnO. Typically the nanorod dimension is, length ~ 1μm, diameter ~ 70-80 nm, and nanorod density ~ $3.2 \times 10^9$ rods/cm$^2$, calculated from the selected area of FE-SEM image. However, there are some nanorods with diameter below 40 nm. Figure 2 shows the top (a) as well as 45° tilt angle (b) views of the nanorod array. Nanorods grow normal to the surface.

The nanorods present a nanopillar shape, and they represent a hexagonal shape from the top. It is observed that at some point that the longitudinal growth is stopped close to 1 μm, which is very unique when Al:ZnO seed layer is used, and has been observed by other authors [9]. The nanorods present a nanopillar shape, and represent a hexagonal shape from the top. It is observed that at some point the longitudinal growth is stopped close to 1 μm, which is very unique when Al:ZnO seed layer is used. The average length of the nanorod is around 1 μm for the current growth time. This is mainly due to the fact that the nucleation centers are first formed on the homostructure of AZO seed layer film. Once the nucleation starts on few grains, forming a hexagonal base as shown in Fig. 2 (c), the density of nanorods remains rather constant after the initial nucleation stage. The TEM image (Fig. 2 (d)) also demonstrates that the ZnO is crystalline for the AZO nanorod shown in the inset of Fig. 2 (d). It is also noted that the energy dispersive spectroscopy (EDS) study of AZO nanorods show Al concentration of about 3 at%, which is slightly more than the Al concentration (~ 2.6%) measured using X-ray photoelectron spectroscopy.

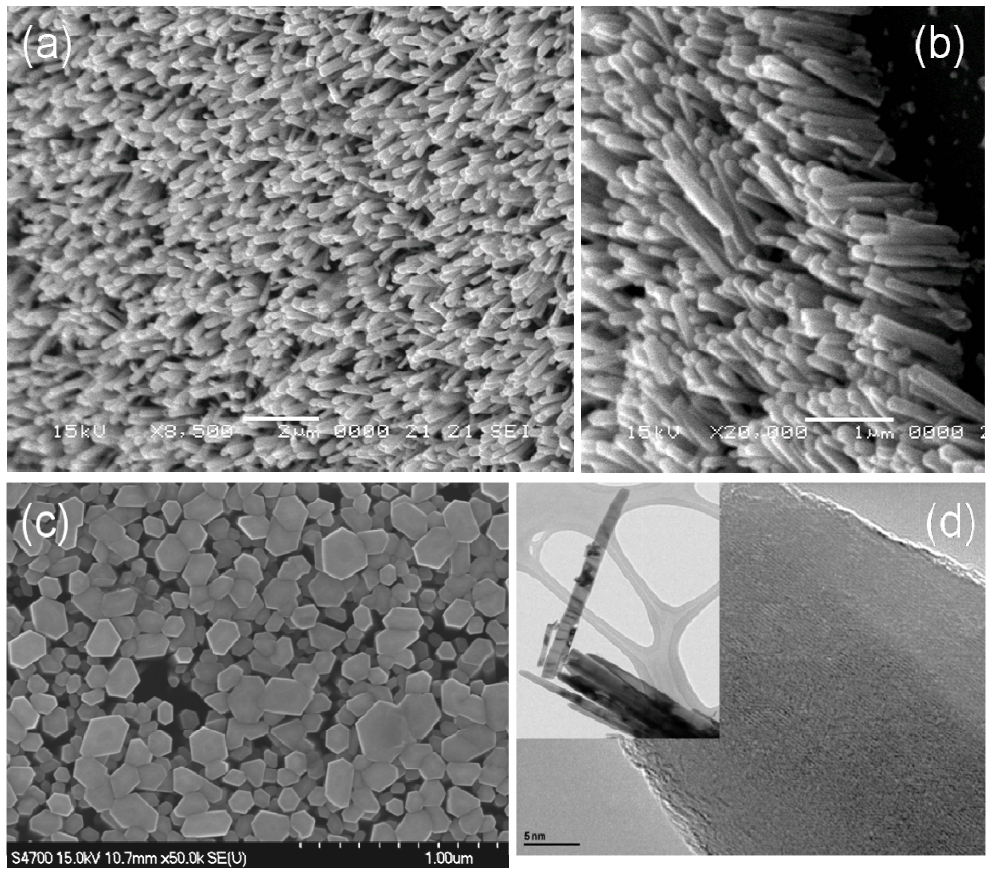

Fig. 2. (a) SEM image of Al:ZnO nanorod array and (b) tilt angle view of the nanorod array, (c) top view of the nanorods, and (d) TEM image of the nanorod in the inset shown. 5wt% aluminum is used in the growth solution.

Figure 3 shows the temperature-dependent resistivity curves measured using a four-probe electrode configuration in order to compare the resistivity of Al:ZnO on glass and p-Si substrates grown at 350 °C. The lower resistivity of the Al:ZnO/glass films grown can be explained due to the Burstein–Moss effect [12] that the Fermi level shifts to the conduction band of a degenerate semiconductor leads to an energy band widening and is consistent with our optical data. The important characteristic is the observation of the metal-like and metal–semiconductor transition behavior. The metal–semiconductor transition (MST) was noticed in the vicinity of 125 K to 130 K for the current films. However, the resistivity continues decreasing down to $T = 100$ K, showing a characteristic up-turn in resistivity behavior at low temperature which is generally found in an impure metal. The film showed a positive temperature coefficient of resistance (TCR) above MST and a negative TCR below it as we reported earlier [13]. The lower resistivity of ZnO/glass than that of ZnO/p-Si can be attributed to the defects which is evident from the higher residual resistivity in ZnO/p-Si. One of the possible reasons is due to the interfacial defects between Si and Al:ZnO film.

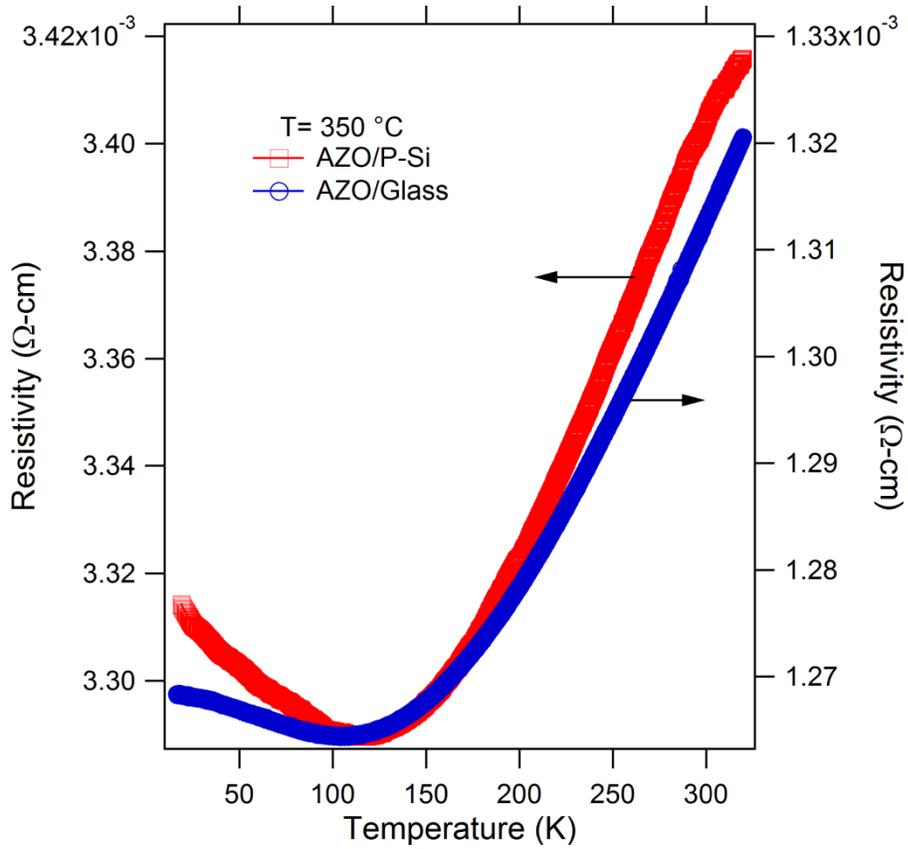

Fig. 3. Temperature dependence of electrical resistivity of Al:ZnO seed layers on glass and P-Si substrates.

Figure 4 shows the optical absorption spectra of AZO films on glass substrates. The absorption edge of AZO/glass shifts to higher wavelength for nanorod arrays grown on AZO seed layer. The absorption edge of AZO/glass film shifts to lower wavelength, indicating an increase in band gap opposite to the nanorod arrays. The transmission of AZO film shows a dramatic reduction from 90% to below 40% for nanorod arrays on AZO seed layer and shown in the inset of Fig. 4. The reason for this reduction can be related to the enhanced diffusion in the visible wavelength region, and trapping of the light in the nanorod arrays. This naturally explains the enhanced absorption in nanorod arrays. In that case the measurement of absorption is not absolute as in presence of diffusion will appear as an additional loss from the ballistic contribution. Hence the total absorption measurement is essential. The antireflecting [14] and scattering effects of highly vertically oriented nanorod arrays with higher surface density as mentioned earlier are the major reasons for reduced transmission for nanorod arrays. Each nanorod acts as a scattering center [15] and enhances the multiple scattering for a certain range of wavelengths due to nanorod arrays. This causes a significant drop in the total transmission from 90% to below 40% as observed in the present case. It is mentioned that the layer is pretty transparent. The scattering of light is very intense when the light falls on the saples with nanorods opposed to AZO/glass seed layer.

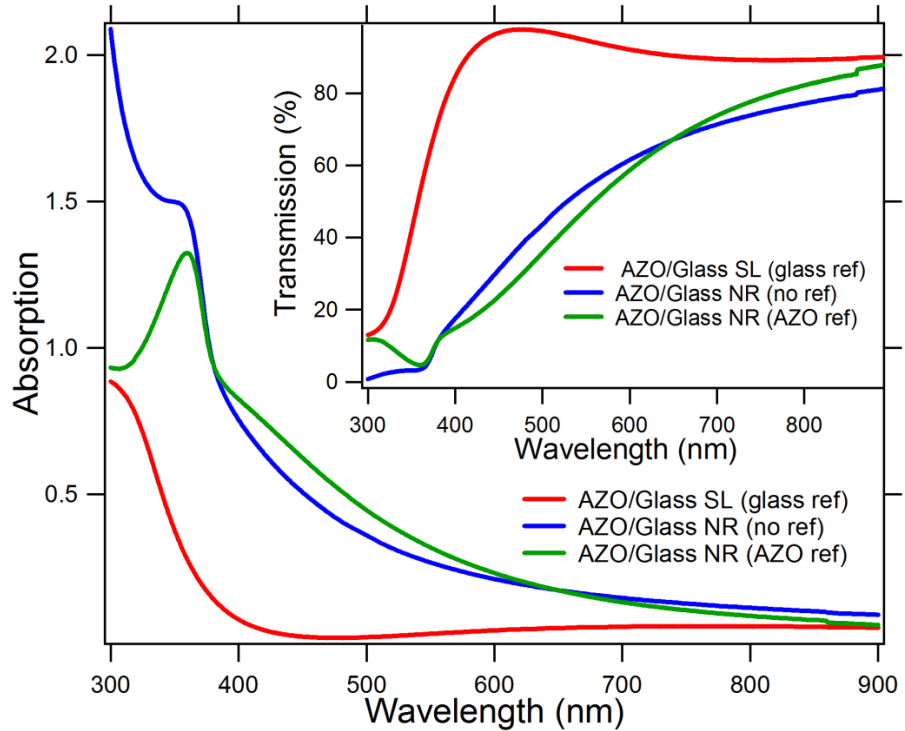

Fig. 4. Absorption spectra of AZO/glass substrate, and nanorods grown on AZO/glass substrates with and without nanorods grown on the surface. AZO/glass is used as a reference substrate. The inset shows the transmission spectra of AZO/glass substrate, and nanorods grown on AZO/glass substrates with and without nanorods.

In a high absorption region, Tauc [16] and Davis and Mott [17] showed that the absorption coefficient and photon energy are related by the following equation.

$$\alpha h\nu = A\,(h\nu - E_g)^n \quad \ldots\ldots\ldots\ldots\ldots(1)$$

In the above equation, $A$ is a constant, $E_g$ is the band gap of the material and $n$ has different values depending on the optical absorption process. It was found that $n = 1/2$ is the best fit to our results, and is characteristic of the direct band absorption without phonon mediation. Figure 5 shows the $(\alpha h\nu)^2$ versus $h\nu$ plot from which the optical band gap was determined. It is very clear that the band gap for the nanorod arrays/AZO/glass is ~ 3.18 eV while AZO/glass seed layer is 3.5 eV. It is noted that the band gap for ZnO film is 3.33 eV. The blue shift of the band gap for nanorod arrays/AZO/glass can be explained due to the effect of quantum confinement while the red shift of the band gap for AZO/glass seed layer is attributed due to Al doping. It is noted that the peak at ~360 nm (~3.44 eV) in samples containing AZO nanorod arrays may be contributed from Al doping. However, the diameter of nanorods prepared in this work is larger than that required for the quantum confinement. One of the possible reasons is that the diameter of some of the nanorods may have smaller size as seen in Fig. 2 (c) (tops of the nanorods, typically less than 40 nm). However, the actual reason for the blue-shift of the band gap at 3.18 eV needs further studies.

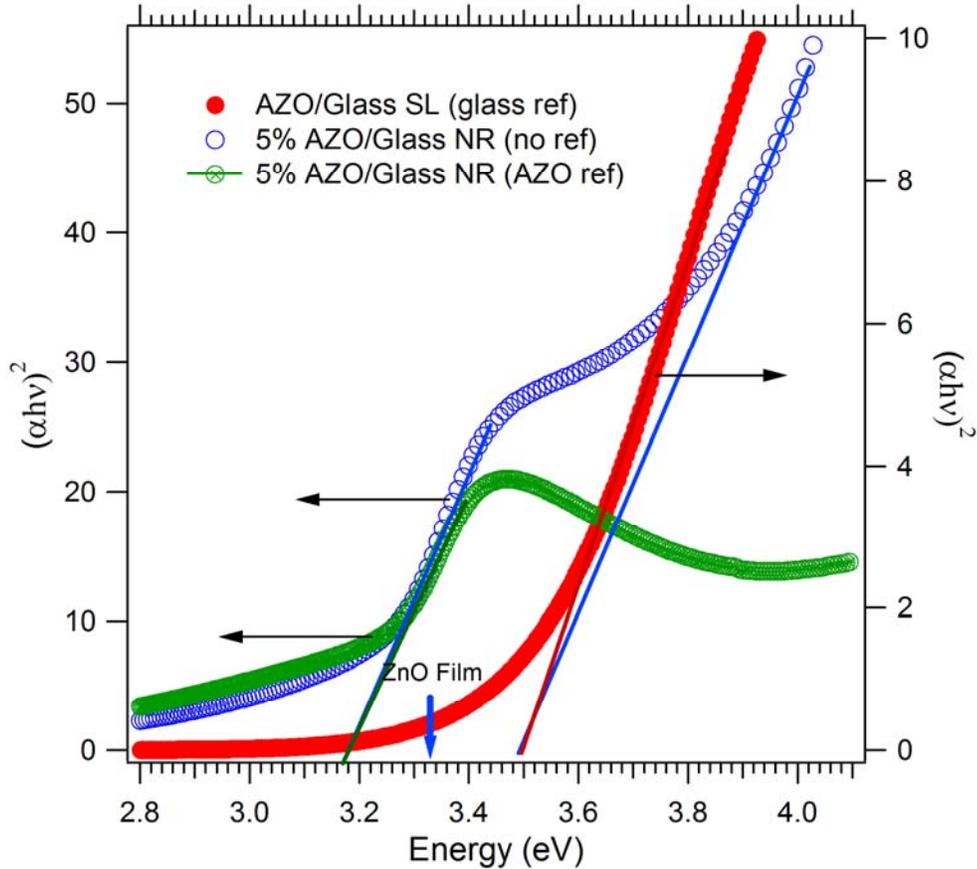

Fig. 5. Optical band gap of AZO/glass with and without nanorod arrays, Glass as well as AZO references were used to distinguish the nanorod contribution.

Al:ZnO film is highly transparent ($T > 90\%$) in the visible region, and that allows the visible light passes through AZO film. For AZO/p-Si the light is primarily absorbed in the underlying p-Si. This generates electron–hole pairs, producing the photocurrent under reverse bias conditions. As the UV photons are mainly absorbed in the AZO layer the photo-generated electrons are drifted towards the positive electrode through the AZO region. This gives rise to the increase in current almost linearly as the reverse bias increases (not shown here). On the other hand, the p-type silicon surface provides a supply of elections which can enter into the AZO film. The effective coupling current flows due to the interchange of the charge between the conduction and valence bands of the silicon by recombination-generation. However, due to linear increase in current with increasing reverse bias, the diode characteristic did not arise. In order to see this effect, the AZO/SiO$_2$/$p$-Si heterojunction was fabricated in which SiO$_2$ works as insulating barrier. This junction was illuminated under the halogen lamp with 1 Watt/cm$^2$ in order to observe the photovoltaic effect. The nanorods were grown on AZO/SiO$_2$/$p$-Si heterojunction. Fig. 6 shows the current density Vs. bias voltage for AZO/SiO$_2$/$p$-Si heterojunction with and without nanorods on the surface of AZO seed layer.

The insulating barrier worked to increase the depletion region between AZO and $P$-Si. It is remarkable to note that photo-current in reduced significantly in sample with nanorods. As discussed earlier, the scattering effects of highly vertically oriented nanorod arrays with higher surface density are responsible for reduced transmission for nanorod arrays and reduced absorption by the seed layer. However, it is anticipated that the similar structure with nanord arrays will enhance the photo-current when dye-sensitized. This is due to the reabsorption of light in the dye due to multiple scattering phenomena by the nanorods.

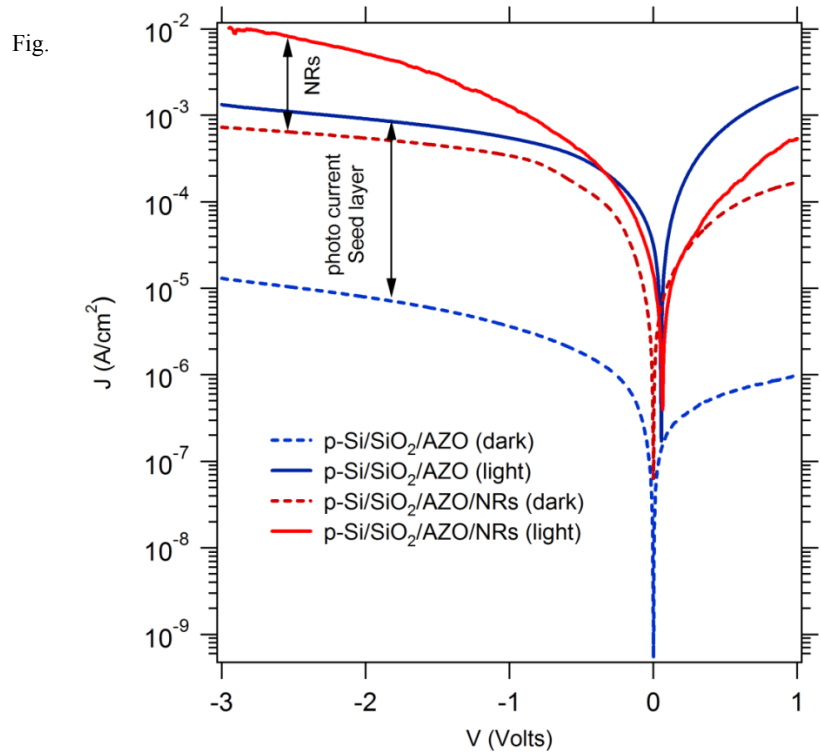

Fig. 6. Current density Vs. Voltage characteristics of the seed layer and seed layer with nanorods. SiO$_2$ barrier layer is used to inhibit channeling of the current.

## 4 CONCLUSIONS

In summary, we report the optical and optoelectronic properties of vertically aligned Al:ZnO nanorod arrays synthesized by the hydrothermal technique at considerably low temperature on a sputtered Al:ZnO seed layer. The morphology demonstrates that the nanorod arrays maintain remarkable alignment along to the c-axis over a large area. The investigation of the optical properties of the seed layers as well as the nanorod arrays show varied band gaps. The optoelectronic properties of nanorod arrays on AZO/$SiO_2$/$p$-Si heterojunction with $SiO_2$ barrier layer have been illustrated. The photocurrent is significantly reduced in nanorod arrays on AZO/$SiO_2$/$p$-Si heterojunction due to multiple scattering phenomena associated with the nanorod arrays. This research may open up venues for various optical and opto-electronic applications where highly aligned nanostructures are desired.

## Acknowledgments

This work is supported by the DoD (CEAND) Grant Number 59025-RT-REP, NSF-CREST (CNBMD) Grant number HRD 1036494, and NSF-RISE-HRD-0931373. The authors are thankful to R.B. Konda for experimental help. All authors have equal contribution.

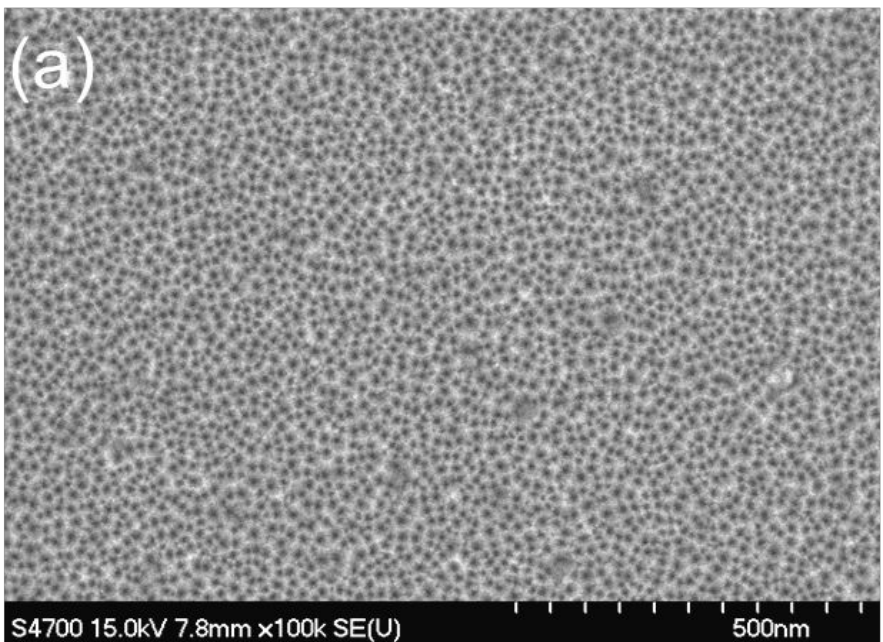

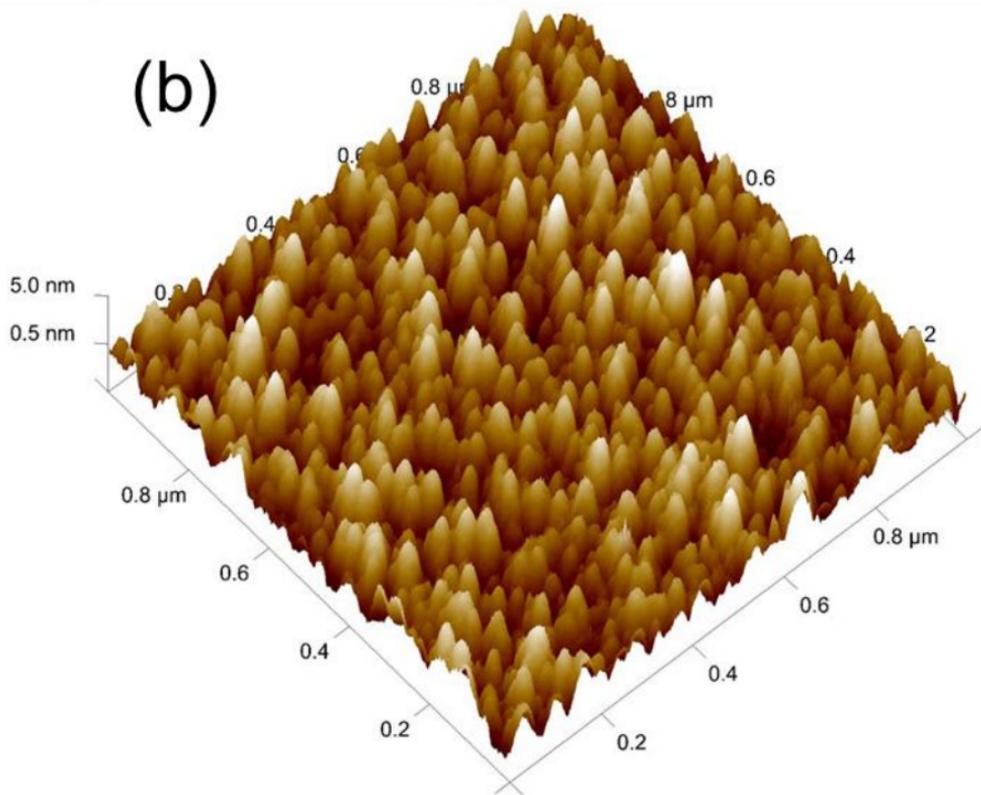

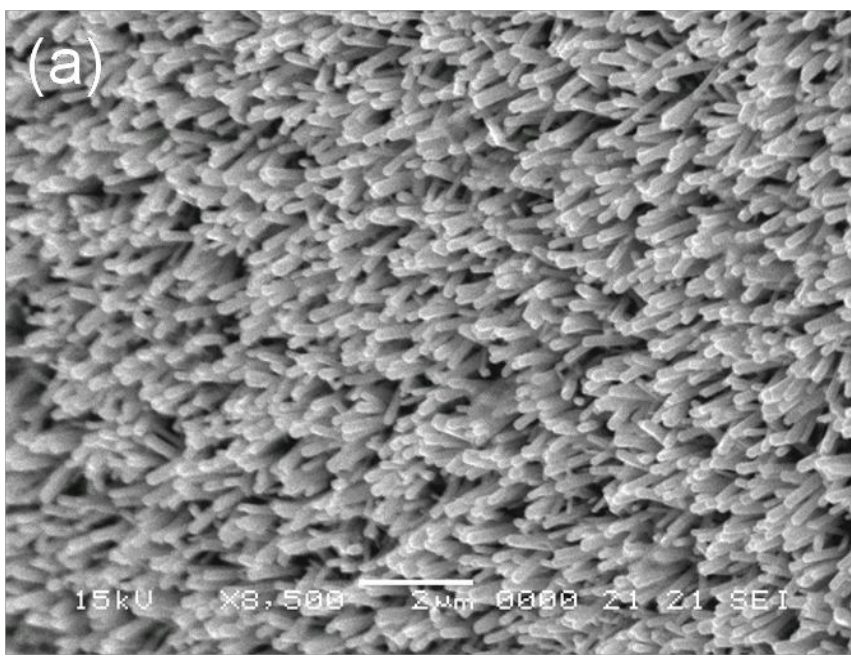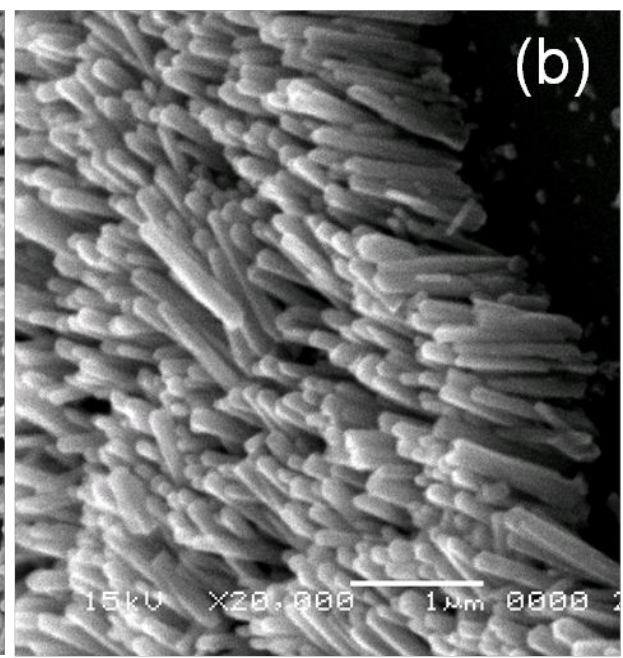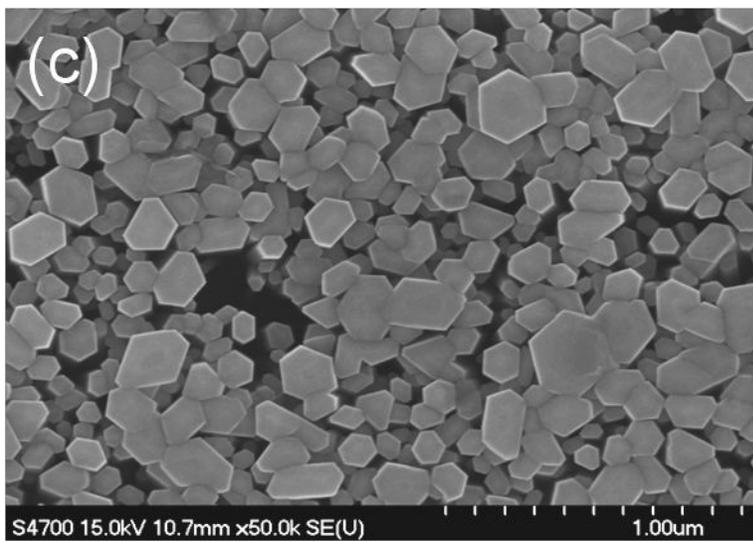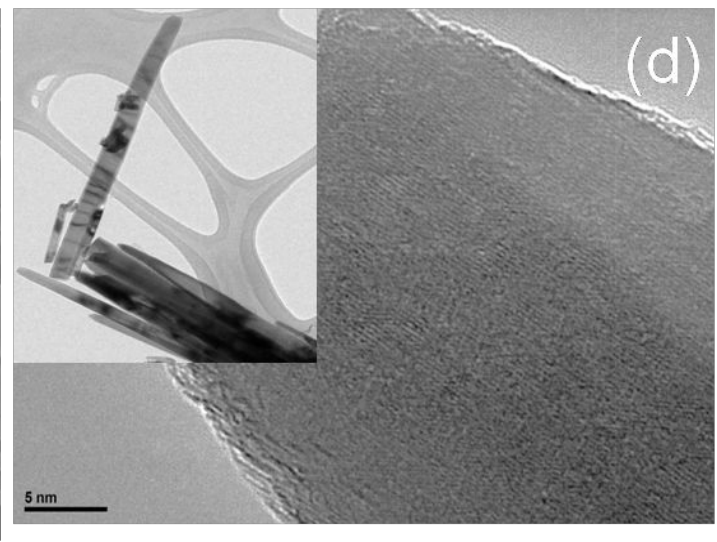

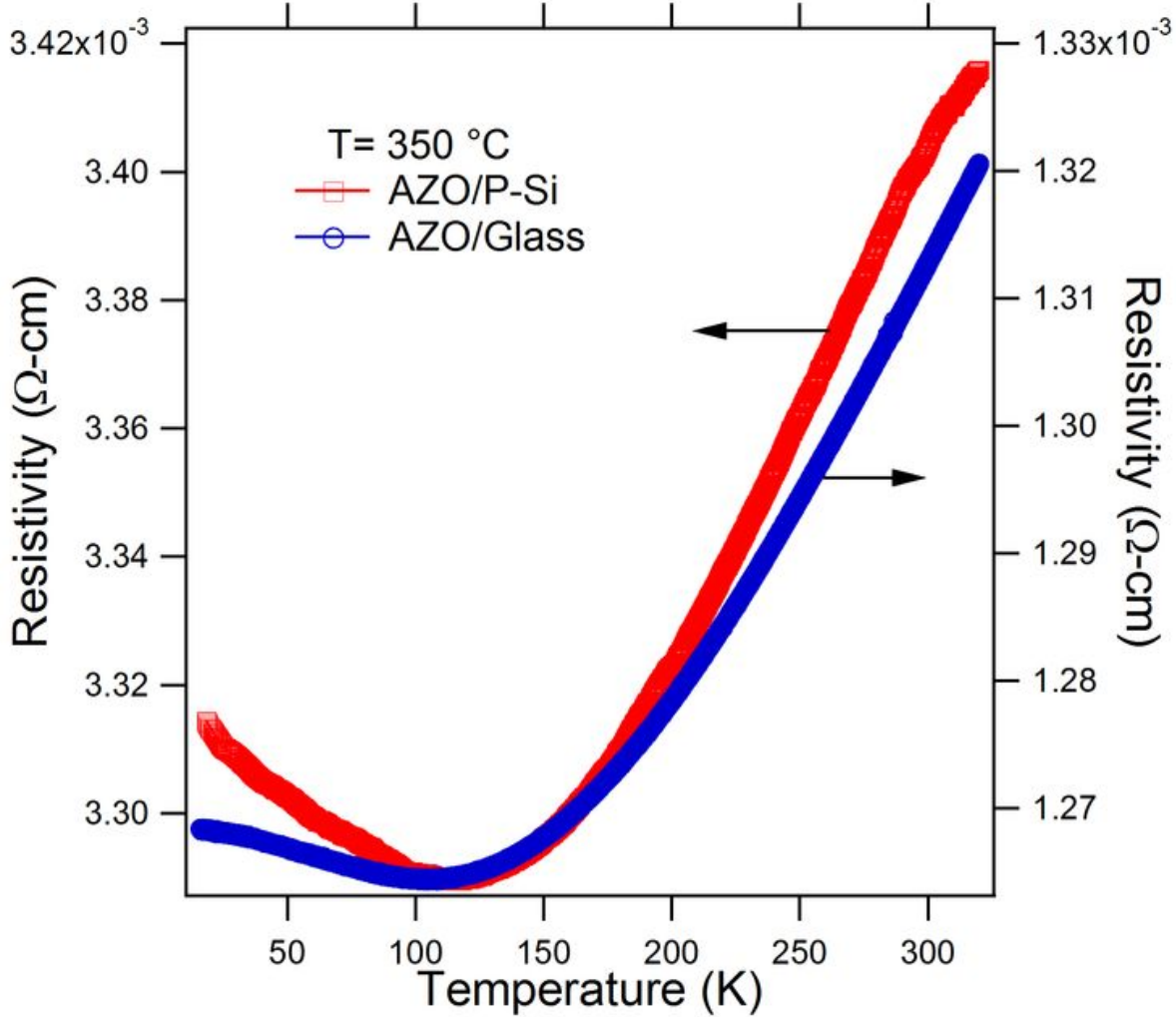

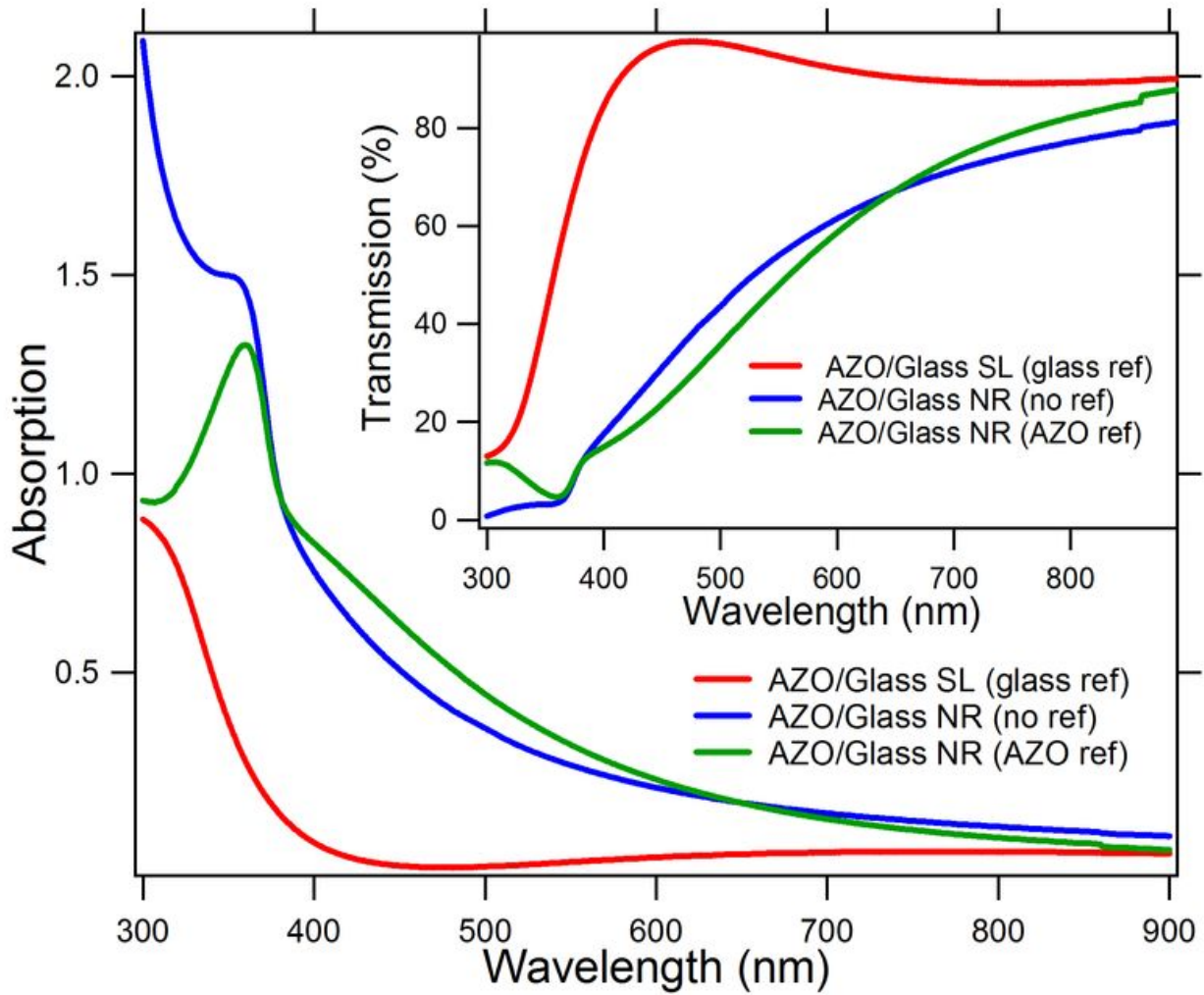

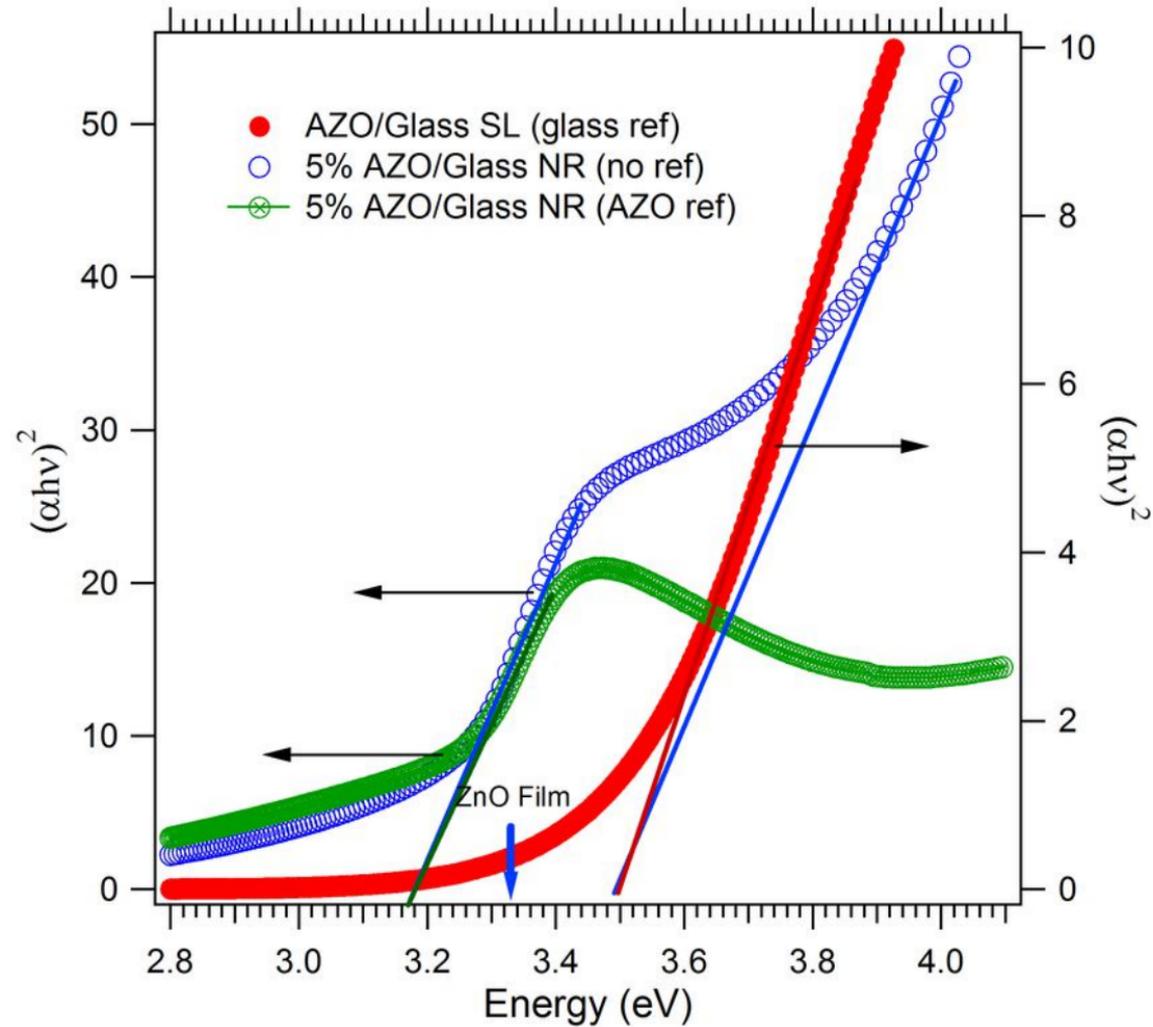

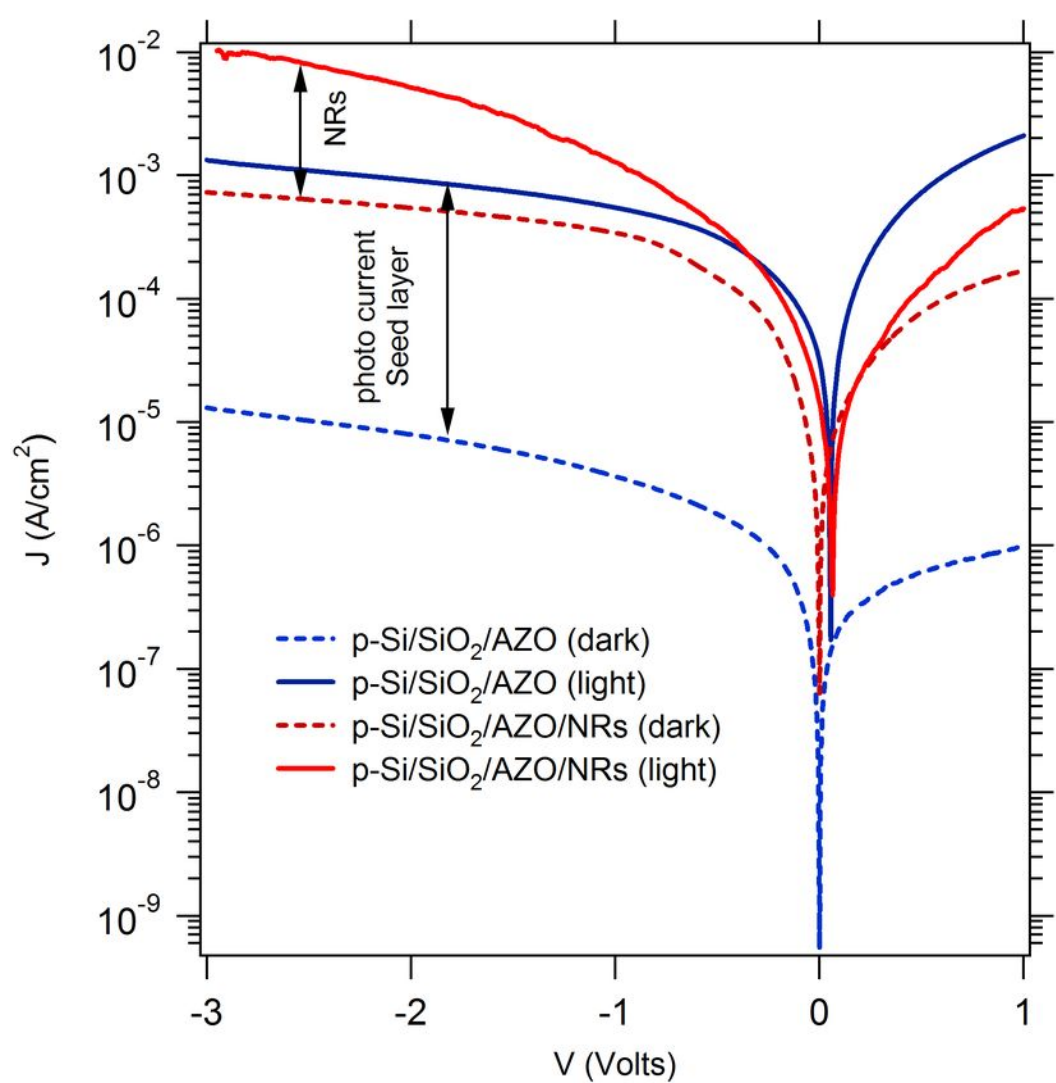